\documentclass[twocolumn,showpacs,preprintnumbers,amsmath,amssymb,aps]{revtex4}

\usepackage{graphicx}
\usepackage{dcolumn}
\usepackage{bm}
\usepackage[T1]{fontenc}
\usepackage{amssymb}
\usepackage{amsbsy}
\usepackage{amsmath}
\usepackage{xspace}
\usepackage{braket}

\newcommand{\mott}{\textsc{2D~MOT}\xspace}
\newcommand{\mottt}{\textsc{3D~MOT}\xspace}
\newcommand{\mot}{\textsc{MOT}\xspace}
\newcommand{\zs}{\textsc{ZS}\xspace}

\usepackage{bm}

\draft
\begin{document}

\title{A compact high-flux source of cold sodium atoms}

\author{G. Lamporesi}
\author{S. Donadello}
\author{S. Serafini}
\author{G. Ferrari}

\email{ferrari@science.unitn.it}

\affiliation{Dipartimento di Fisica, Universit\`a di Trento and\\
INO-CNR BEC Center -- I-38123 Povo, Italy, EU}

\date{\today}

\begin{abstract}
We present a compact source of cold sodium atoms suitable for the production of quantum degenerate gases and versatile for a multi-species experiment. The magnetic field produced by permanent magnets allows to simultaneously realize a Zeeman slower and a two-dimensional MOT within an order of magnitude smaller length than standard sodium sources. We achieve an atomic flux exceeding $4\times10^9$~atoms/s loaded in a MOT, with a most probable longitudinal velocity of 20~m/s, and a brightness larger than $2.5\times10^{12}$~atoms/s/sr. This atomic source allowed us to produce a pure BEC with more than $10^7$~atoms and a background pressure limited lifetime of 5 minutes.
\end{abstract}



\maketitle

\section{INTRODUCTION}    

Laser cooling techniques developed in the last thirty years revolutionized atomic physics allowing to cool neutral atoms down to micro and nanokelvin temperatures, at which ultimate control of each atomic degree of freedom becomes possible. Nowadays cold atoms are employed routinely
in metrological applications \cite{Chu02}, in the realization of inertial sensors (\cite{deAngelis09} and Refs. therein) and to operate state-of-the-art atomic clocks \cite{Katori11}. At these temperatures dilute gases can also reach quantum degeneracy offering the possibility to directly manipulate quantum degenerate systems with extreme precision and control (see \cite{Anglin02} and Refs. therein).\

The growing interest in cold atoms led to the development of specific atomic sources. Sources capable of high fluxes have been developed
for many atomic species addressing issues of compactness, atomic yield, and ease of use. In fact the availability of essential experimental
tools as the atomic source, in many circumstances proved to drive the choice of the atomic species to study and, eventually, the
physical domain to address. So far high fluxes of cold atoms are obtained mainly with two classes of atomic sources. In the first one a flux of hot atoms from an oven is slowed down by means of dissipative light forces and inhomogeneous magnetic fields \cite{Phillips82}. This technique is applied to a wide
class of atomic species, such as alkali metals, alkali earths, rare earth elements and noble gases. The second class of atomic sources is based
on the trapping and cooling of atoms directly from vapor pressure. This simplifies the experimental setup, but the performances are significant
only for some medium-heavy alkali atoms such as potassium \cite{Catani06}, rubidium \cite{Dieckmann98} and cesium \cite{Yu94}. \

In this article we present a novel type of compact atomic source delivering cold sodium atoms with state-of-the-art fluxes \cite{vanderStam07}. This sodium source has a novel design that combines high-flux performances with the compactness and simplicity of the experimental setup. It is
based on a thermal sodium atomic beam coming out from an oven that is slowed down and two-dimensionally trapped. An additional laser beam aligned along the non-trapped direction pushes the atoms hence obtaining a collimated and slow atomic beam that is finally recaptured and cooled in a three-dimensional magneto-optical trap (\mottt) in a nearby ultra-high vacuum (UHV) region.\

Sodium is a valuable option if one needs to produce very large Bose-Einstein condensates, thanks to its small three-body recombination rate and convenient scattering length. In the last years interest in producing cold samples of sodium comes also from the ultra-cold molecules field community \cite{Ni08}; together with KCs, NaK  seems \cite{Wu12} in fact to be an excellent pair of alkali atoms for the production of stable systems \cite{Zuchowski10} of ground state molecules with a dipole moment of a few debyes \cite{Gerdes11} that therefore represents a promising system for accessing a new domain of cold atomic/molecular physics passing from contact interaction to long-range, anisotropic interactions.\

With the exception of a few experiments in which a small number of atoms is needed \cite{Mimoun10}, standard cold sodium experiments employ a long Zeeman slower (\zs) \cite{vanderStam07} to efficiently slow down a large number of atoms coming out from an oven. Sodium 2D MOTs were already demonstrated as atom funnels \cite{Riis90} to increase the brightness of traditional \zs. \

The work reported in \cite{Tiecke09} introduced an innovative approach in the capture of light atoms (lithium in that case) from a thermal distribution directly in a \mott placed in the vicinity of the oven. We demonstrated this approach for sodium atoms and we showed how a shrewd choice of magnetic fields and the addition of a further laser beam results in higher yields of the atomic source. The cold sodium atom source thus realized represents a compact and convenient alternative to the classic one for loading a large number of atoms in a \mottt. \

\begin{figure*}[!hbt]
\centering
\includegraphics[width=2.07\columnwidth]{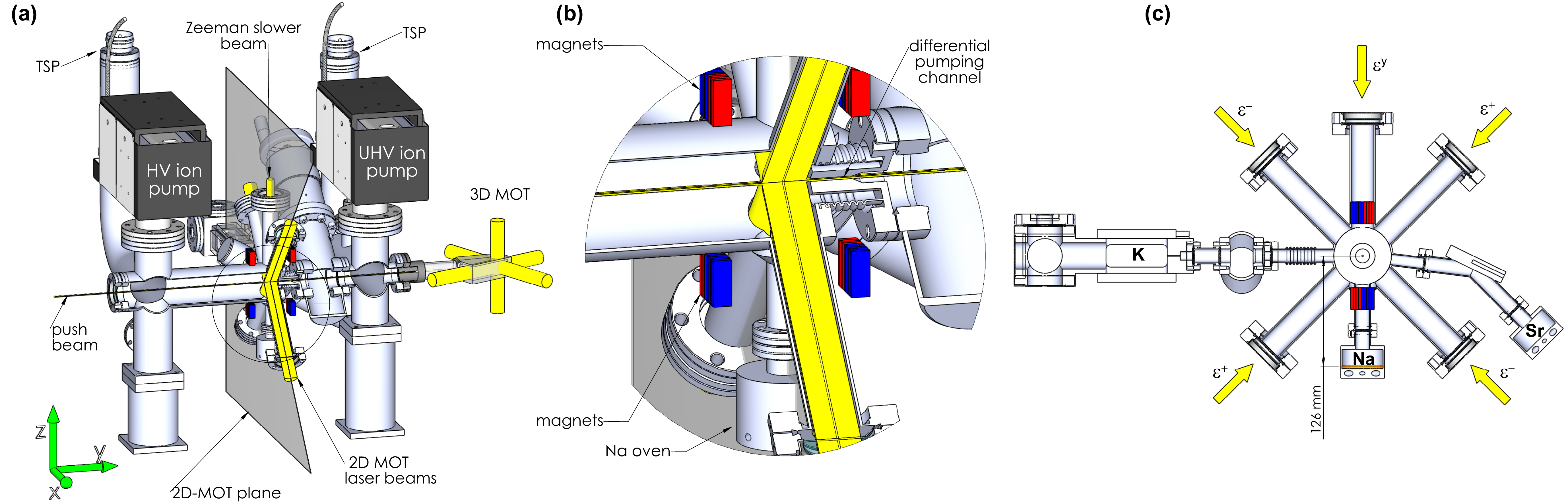}
\caption{(Color online) (a) 3D view of the vacuum system. HV region on the left side contains the atomic source and the optical access for the pre-cooling stage. The differential pumping channel connects this to the UHV region where the experiment is performed in a clean environment. Light beams (yellow) and magnets (red-blue) are shown. (b) Magnification of the compact slowing/cooling region. (c) 2D view of the pre-cooling plane showing atomic sources and beams configuration.}
\label{fig:VacuumAll}
\end{figure*}

In the following sections the reader will find a description of the atomic source working principle with details of the the vacuum, optical and magnetic setup (Sec.~\ref{APPARATUS}), a set of the most meaningful characterization curves (Sec.~\ref{SYSTEM CHARACTERIZATION}) that were experimentally recorded on the apparatus, and a simulation (Sec.~\ref{SIMULATION}) of the atomic trajectory compared with the final performances obtained. \

\section{APPARATUS DESCRIPTION}   
\label{APPARATUS}
A sketch of the experimental setup is given in Fig.~\ref{fig:VacuumAll}. Sodium atoms evaporated from an oven, heated to 210$^{\circ}$C, propagate in the vertical direction towards the \mott region. Along their trajectory a red-detuned counter-propagating beam, in combination with a magnetic field increasing with $z$, slows a large velocity class of atoms below the capture velocity of a \mott located just 12 cm above the oven. The \mott is in a vertical plane and atoms are free to move along a horizontal transfer axis. A beam aligned on that axis pushes the cold captured atoms through a differential pumping channel towards a cleaner environment where a \mottt collects the incoming atomic flux. \

The main features and some technical details of the experimental apparatus will be provided in the following. \

\subsection{Vacuum system}    

Standard cold atoms experiments typically need to suppress background vapor pressure to extend the lifetime of the sample on the minute timescale. At the same time large number of cold atoms collectable in the final trap is desirable. These two requirements are difficult to be simultaneously achieved in the same vacuum chamber, hence in our design the vacuum system is divided in two regions (HV and UHV chambers) connected by a narrow channel which ensures differential pumping. The chamber on the left side of Fig.~\ref{fig:VacuumAll}(a) is mainly devoted to the atomic source with a pre-cooling stage, while the one on the right is designed for trapping cold atoms and studying them in UHV conditions.  \

The differential pumping channel, with diameter $d$=2.0~mm  and length $l$=22.8~mm, in the long tube approximation has a conductance of $4.3\times 10^{-2}$~l/s. Assuming the nominal pumping velocity of the pump in the UHV region, from the conservation of the mass flow we can obtain a differential pressure between the HV and UHV chambers up to $10^3$. \

The geometry of the \mott chamber is inspired to the one described in \cite{Tiecke09}. In a vertical plane several \textsc{AISI 316L} stainless steel tubes cross allowing for atomic and optical access for pre-cooling. 5 grams of metallic sodium are held in a crucible positioned in the plane 126~mm below the \mott region and connected through a CF16 flange. The oven is heated up to temperatures of the order of 210$^{\circ}$C, higher than the melting temperature of sodium ($T_{\text{m}}$=97.8$^{\circ}$C), obtaining a flux of fast moving atoms. Our oven operating temperature is 100$^{\circ}$C less than the typical temperatures of sodium ovens employed in combination with a \zs \cite{vanderStam07}. This is a good point for the reliable operation on the long term and for ensuring a high quality of vacuum already in the HV region. \

Atoms collected in the \mott and pushed with an on-axis quasi-resonant beam cross the differential pumping channel and enter the UHV chamber at the end of which a quartz cell hosts a \mottt and other atomic traps for future experiments.  \

There are many advantages in using a transversely loaded \mott instead of atomic sources with a coaxial loading. Atoms from the oven cannot pass directly to the UHV chamber through the differential pumping channel, therefore no extra stages, such as mechanical shutters, are needed in the vacuum chamber to reduce background hot atoms in the recapture region. The transfer of atoms to the \mottt can be optically modulated by switching off the push and \mott beams; in our case no atoms are detected in the \mottt in absence of these beams.\

In addition to the simplification of the apparatus, our approach offers the possibility to simultaneously deal with more atomic species. Thanks to its radial symmetry the \mott can be transversely loaded from different sources (see Fig.~\ref{fig:VacuumAll}(c)). Our setup is already set for cooling also potassium from a vapor-cell \mott. Also a strontium oven is present for future developments. The cooling lights for the operation of different atomic species 2D MOTs are mixed with dichroic mirrors.\

\subsection{Laser system}   

\begin{figure}[b!]
\centering
\includegraphics[width=1\columnwidth]{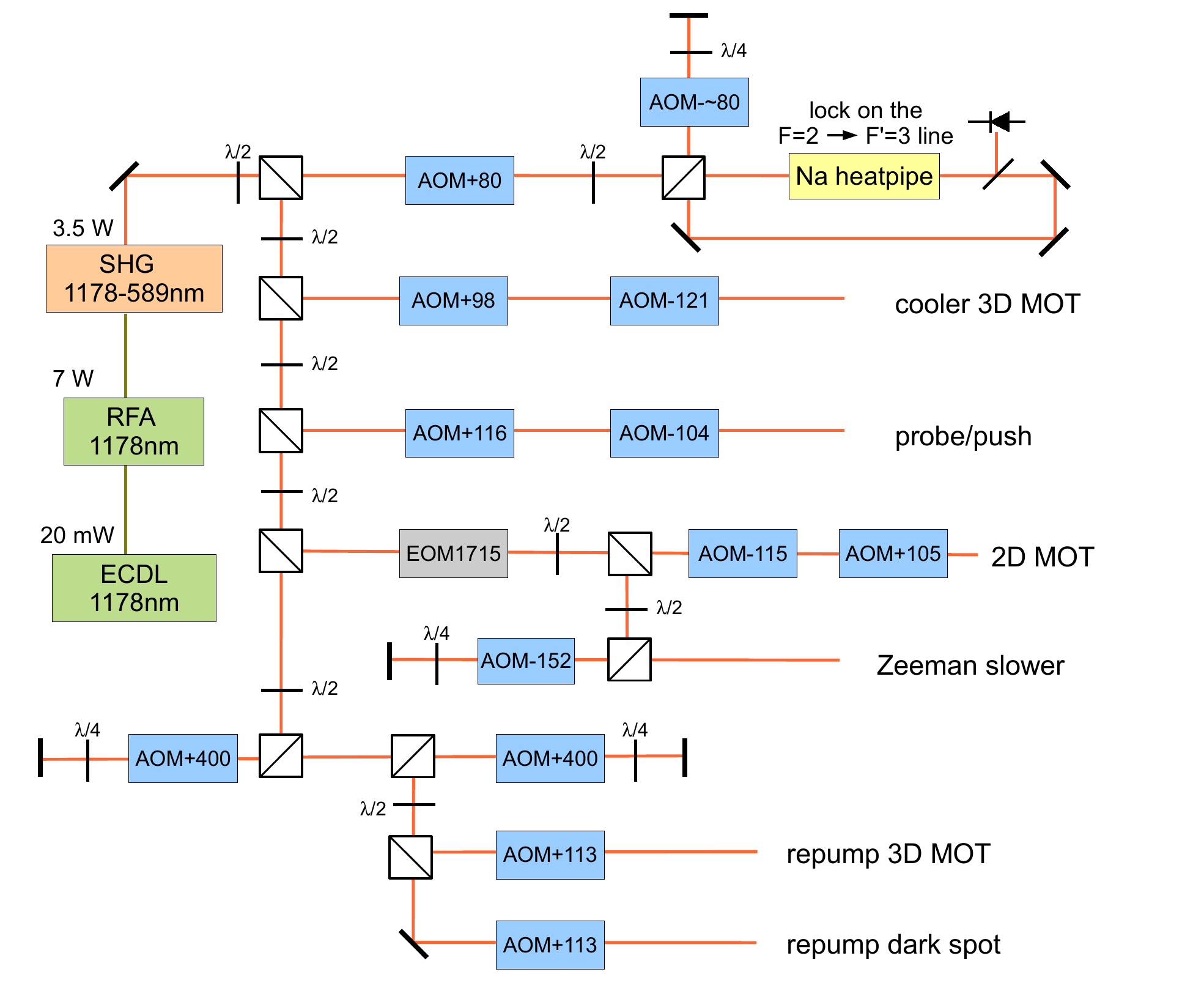}
\caption{(Color online) Sketch of the optical setup used for producing all the light beams needed to cool sodium atoms. Reported power values indicate the power output of the devices, without considering further power losses. Signed numbers in the blue AOM boxes report the chosen AOM order ($\pm$1) and its driving frequency in MHz.}
\label{fig:LaserSource}
\end{figure}

Laser cooling of sodium can be achieved by using cooling light slightly red-detuned from its $\Ket{F=2}\rightarrow\Ket{F'=3}$ transition on the $D_2$ line and repumping light tuned close to the $\Ket{F=1}\rightarrow\Ket{F'=2}$. $D_2$ line for sodium lies at 589 nm, hence not directly accessible with diode lasers. Dye lasers can be employed at this wavelength, but are usually large, expensive, and involved to operate on a long period. On the other hand, the infrared region of the spectrum around 1200 nm, twice the wavelength of sodium $D_2$ line, has recently become accessible with quantum dots technology \cite{Nevsky08}. \

Fig.~\ref{fig:LaserSource} illustrates our approach that consists in frequency doubling a MOPA system delivering 1178 nm light. The master oscillator is a diode laser based on InAs quantum dots on GaAs substrate (\textsc{Innolume GC-1178-TO-200}), with single transverse mode and an anti-reflection coating on the output facet and it delivers about 20 mW. A Raman fiber amplifier (\textsc{MPB RFA-SF-series}) pumped with an Ytterbium fiber laser provides up to 8 W output power on a single transverse mode, when injected with 9 mW,  while maintaining polarization and spectral properties of the incoming beam. \

The frequency of the amplified infrared light is doubled through a resonant frequency doubling unit, based on a 15~mm long LiB$_3$O$_5$ non-linear crystal. The birefringent phase matching is achieved with temperature tuning at 45$^{\circ}$C. The resonant bow-tie cavity has a round-trip length of 300~mm, a finesse of 150, and it is stabilized by means of polarization spectroscopy \cite{Hansch80}. \

When operating the RFA at 7 W we obtain 3.5 W of light, stabilized at 589 nm. The emission linewidth is about 20 times narrower than the natural linewidth for the $D_2$ transition of sodium. A stable and reliable frequency reference is obtained directly from the $D_2$ transition of sodium, with FM saturated absorption spectroscopy performed on a sodium heat pipe. \

As sketched in Fig.~\ref{fig:LaserSource}, repumping light is produced in two ways: an EOM tuned at 1.713 GHz provides two frequency sidebands, one of which is resonant with the repumping transition, and we employ it for the operation of \zs and \mott. In addition a series of two 350 MHz AOMs in double pass and a single pass one shift the cooling light frequency again by 1.713 GHz. This solution instead, is used to independently control repumping light in the final \mottt region and for a state selective imaging. \

Polarization maintaining optical fibers are used to deliver light on the experimental apparatus providing good quality TEM$_{00}$ mode and decoupling the laser sources from the optical table hosting the vacuum system. All the beams have a diameter of 25 mm with the exception of the push beam that is focussed to a waist of 320 $\mu$m in the differential pumping channel. The total power after the optical fibers is around 600 mW.\

\subsection{Magnetic field sources}    
\label{Magnetic field sources}

Neodymium bar magnets (\textsc{Eclipse N750-RB}) generate the permanent field that, at the same time, is used for slowing down hot atoms coming from the oven in a \zs-like configuration and for trapping and cooling them in a \mott. These magnets are relatively small, ($10\times25\times3$)~mm$^3$, favoring the development of compact atomic sources, and provide a stable magnetic field when operated below 100$^{\circ}$C. The magnetization and the corresponding point-like dipole for a single magnet, resulting from our measurements, are $M$=$(8.7 \pm 0.1)\times10^5$ A$\,$m$^{-1}$ and $m$=$(0.65 \pm 0.01)$ A$\,$m$^2$, in good agreement with the value reported in \cite{Tiecke09} for the same magnets used for a lithium \mott. \

\begin{figure}[t!]
\centering
\includegraphics[width=1\columnwidth]{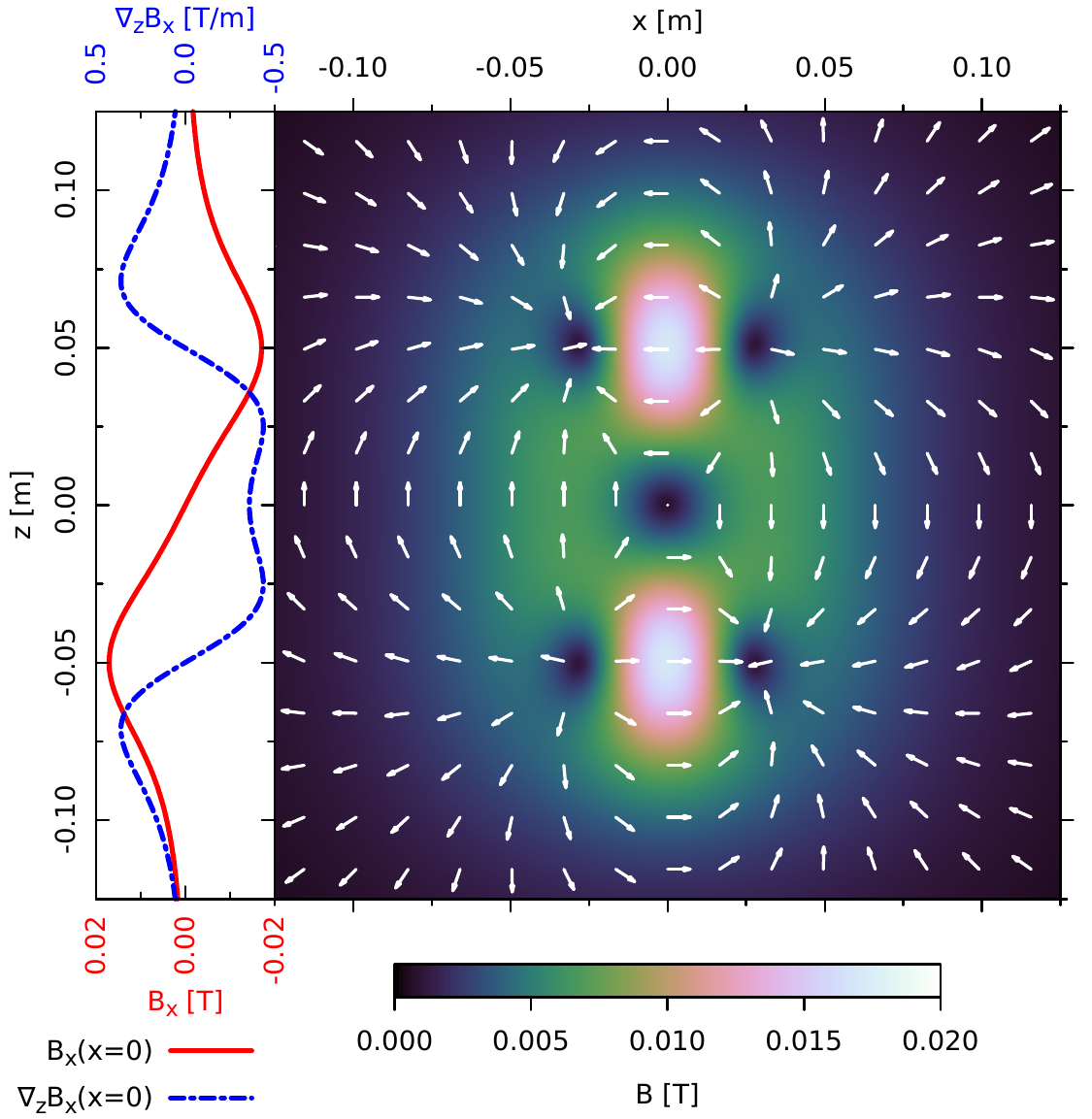}
\caption{(Color online) Representation of the field generated by the magnets in the \mott central vertical plane ($y$=0) orthogonal to the chamber axis. The color scale is a function of the magnitude of the field, while the vectors represent its direction. The four magnet stacks are located in $(x,y,z)$=$(0,\pm37,\pm49)$~mm, the two on top being oriented along $-x$ and oppositely the other two. The plot on the left shows magnetic field and gradient along the central ($x$=$y$=0) vertical axis of the \mott, along which the \zs acts. The field along $y$ axis is zero.}
\label{fig:MagnetsField}
\end{figure}

Given the spatial constraints of the vacuum apparatus, we found that the optimal  magnets arrangement is achieved by placing four equal stacks of nine magnets each at the corners of a hypothetical vertical rectangle centered on the \mott axis (see Fig.~\ref{fig:VacuumAll}). To get a quadrupolar field the upper dipoles are aligned perpendicularly to the plane and the lower ones in the opposite direction. In comparison with \cite{Tiecke09} we replaced each in-plane magnetic stack with a pair of two stacks equally distant from the cooling plane in order to leave access to the main axis of the vacuum system allowing for multi-species atomic in-flow and for the \zs beam optical access. Furthermore the plane of the magnets is rotated by 90$^{\circ}$ with respect to the \mott plane, leaving the operation of the \mott unaltered, but ensuring about a 3-fold increase of the peak magnetic field along the \zs trajectory. \

The magnet stacks are fixed along the chamber axis near the vertical tubes of the \zs and the oven, with centers at a vertical distance of 98~mm and a horizontal one of 75~mm. \

The total calculated magnetic field for the final configuration of the \mott in the $xz$ plane (see notation in Fig.~\ref{fig:VacuumAll}) is reported in Fig.~\ref{fig:MagnetsField}. On the central vertical axis the field is directed along $x$. Its magnitude as a function of $z$ is also reported in Fig.~\ref{fig:MagnetsField}. The expected vertical and horizontal gradient in the middle is 0.36 T$\,$m, while the largest magnitude of the field is $1.71 \times 10^{-2}$ T along the vertical axis and $6.6 \times 10^{-3}$ T along the horizontal one. No field is present along the \mott axis ($y$). \

Since we cannot turn off the magnets field once the \mottt loading is completed, we evaluated the magnitude of the field in the \mottt region. It results being at most $2\times10^{-7}$ T with a maximum gradient of $3\times10^{-5}$ T$\,$m$^{-1}$, both much smaller than the typical values for MOT operation. \

The main drawback of our magnetic configuration consists in the fact that at most only half of the power of the \zs beam has a well-defined circular polarization, as desired for an optimal operation. Given the field along $x$ and the \zs beam propagating along $z$, in fact, one can maximize the amount of circular polarization by polarizing the \zs slower beam along $y$. \

\section{SYSTEM CHARACTERIZATION}   
\label{SYSTEM CHARACTERIZATION}

\begin{figure*}[t!]
\centering
\includegraphics[width=2.07\columnwidth]{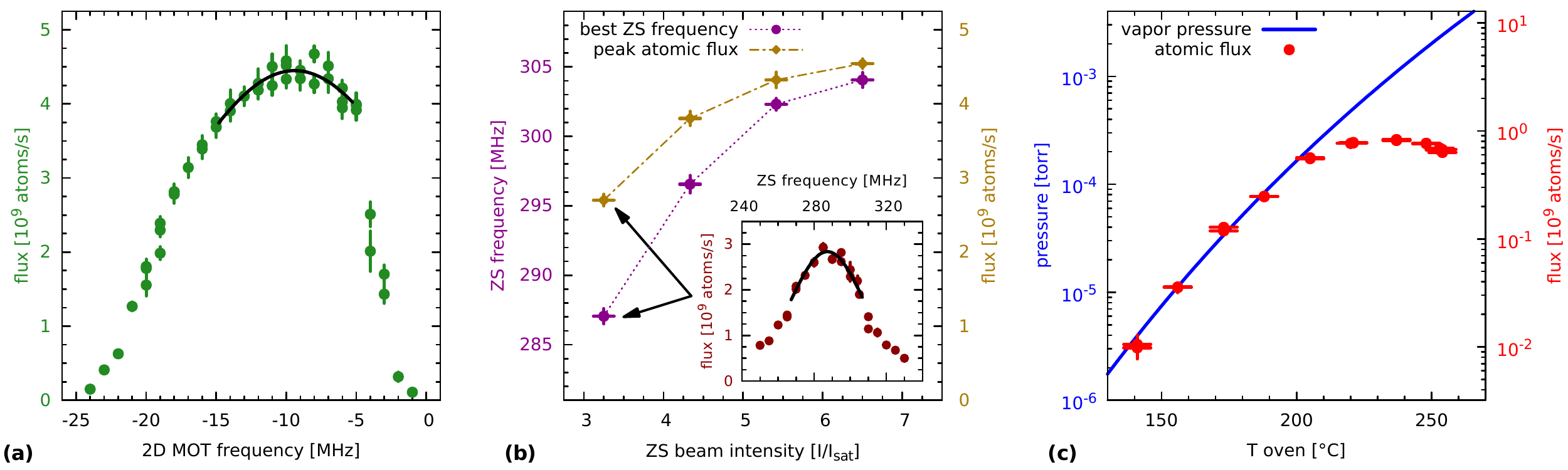}
\caption{(Color online) (a) Measured flux of atoms (green circles) captured in the \mottt as a function of the frequency detuning of the \mott beams. Central data are fit to a Gaussian (black line) to find the best frequency. (b) Best \zs frequency (red) and maximum flux recorded (yellow) for different \zs intensities. The inset shows a typical dependence of atomic flux on the \zs frequency at fixed intensity. (c) Comparison between the vapor pressure of sodium (blue line) and the \mottt loading rate (red circles) as a function of the oven temperature. For this dataset about 8 times smaller repumping power was used in \mott and \zs, explaining the smaller maximum flux achieved.}
\label{fig:Characterization}
\end{figure*}

The atomic source was optimized by measuring the loading rate of the \mottt as a function of several parameters of \mott, \zs, push beams and the oven temperature. For very large atom number in the \mottt rescattered light from cold atoms and excited state collisions limit the maximum atomic density achievable. We circumvented this issue by using a dark-spot \mot scheme \cite{Ketterle93}. \

Intensity and detuning of each beam are respectively reported in sodium units of saturation intensity for $\sigma^\pm$ polarized light $I_{\text{sat}}$=6.26~mW/cm$^2$ and natural linewidth for the $D_2$ line $\Gamma$=$2\pi\cdot 9.79$~MHz. \

Table~\ref{tab:parameters} summarizes the system parameters, experimentally found, that provide the best performances of the apparatus. Reported intensity values for MOTs are meant as total intensities summing over all MOT beams, whereas \zs intensity corresponds to the $\sigma^+$ component at the cooling frequency (the beam contains 50\% power on the cooling, 25\% on the repumping sideband and has a linear polarization along $y$).

\begin{table}[h]
\caption{Set of frequency detuning (from the $\Ket{F=2}\rightarrow\Ket{F'=3}$ transition) and intensity for each beam in the atomic source.}
\label{tab:parameters}
\centering
\footnotesize
\begin{tabular}{llrlll}
\toprule
$\mott \text{cool}$     &\hspace{0.7 cm}    &$-10$ MHz    &($-\Gamma$) &\hspace{0.7 cm}    &3.6 $I_{\text{sat}}$ \\
$\mott \text{rep}$      &                   &$+1713$ MHz  &           &                   &1.8 $I_{\text{sat}}$ \\
$\zs$                   &                   &$-304$ MHz   &($-31\,\Gamma$) &                   &6.5 $I_{\text{sat}}$  \\
$\mottt \text{cool}$    &                   &$-33$ MHz    &($-3.4\,\Gamma$)&                   &2.2 $I_{\text{sat}}$  \\
$\mottt \text{rep}$     &                   &$+1718$ MHz  &&                   &0.5 $I_{\text{sat}}$  \\
$\text{push}$           &                   &$+12$ MHz    &($+1.2\,\Gamma$) &                   &11 $I_{\text{sat}}$  \\
\hline
\hline
\end{tabular}

\end{table}

The sensitivity of the system performances to any parameter was explored by varying one of them at a time and the most significant are reported in the following. \

Fig.~\ref{fig:Characterization}(a) shows the atomic flux as a function of the \mott cooling frequency. One can see that the best detuning is around $-1\,\Gamma$ and that $\pm 0.5\,\Gamma$ away from that detuning the efficiency drops by 10\%. This result can be compared with \mott systems of other atomic species. An optimal detuning of $-1.7\,\Gamma$ was found for $^{87}$Rb \cite{Dieckmann98}, that is characterized by a much better defined hyperfine structure; larger detunings were used for $^{39}$K and $^{41}$K where the repumping light also has a cooling effect \cite{Catani06}; optimum detuning was even larger ($-8\,\Gamma$) in \cite{Tiecke09} for $^6$Li whose hyperfine structure is not only narrow, but also inverted. \

The optimization of the \zs parameters is reported in Fig.~\ref{fig:Characterization}(b). \zs frequency was scanned for a few given intensity values. We noticed that the optimum frequency, maximizing the atomic flux, changes with \zs intensity and also that the maximum flux detected for each intensity value varies as power is increased, starting to show a saturating behavior.

Atoms trapped in the \mott are pushed towards the \mottt by using a beam focussed in the differential pumping channel (waist $w$=320 $\mu$m) with 110 $\mu$W power. We observe best transfer efficiency when using a blue-detuning of 12 MHz. No repumping light is present on this beam. \

The optimization of the sodium oven temperature keeping all other parameters fixed is reported in Fig.~\ref{fig:Characterization}(c). We see that for oven temperatures lower than 160$^{\circ}$C the atomic flux is negligible. Then the loading rate increases with the oven temperature up to about 240$^{\circ}$C. For higher temperatures the loading rate decreases. The loading rate is compared with the theoretical vapor pressure predicted for liquid sodium \cite{Alcock84}. We observe that the two parameters are directly proportional for temperatures lower than 200$^{\circ}$C: the initial growth of the atomic flux can be explained with the rise of the oven yield, proportional to the sodium vapor pressure until it reaches $2\times10^{-4}$~torr. At higher temperatures the loading rate flattens and decreases, likely because of collisional losses induced by the hot beam hitting the trapped atoms in the \mott. \

\begin{figure}[b]
\centering
\includegraphics[width=1\columnwidth]{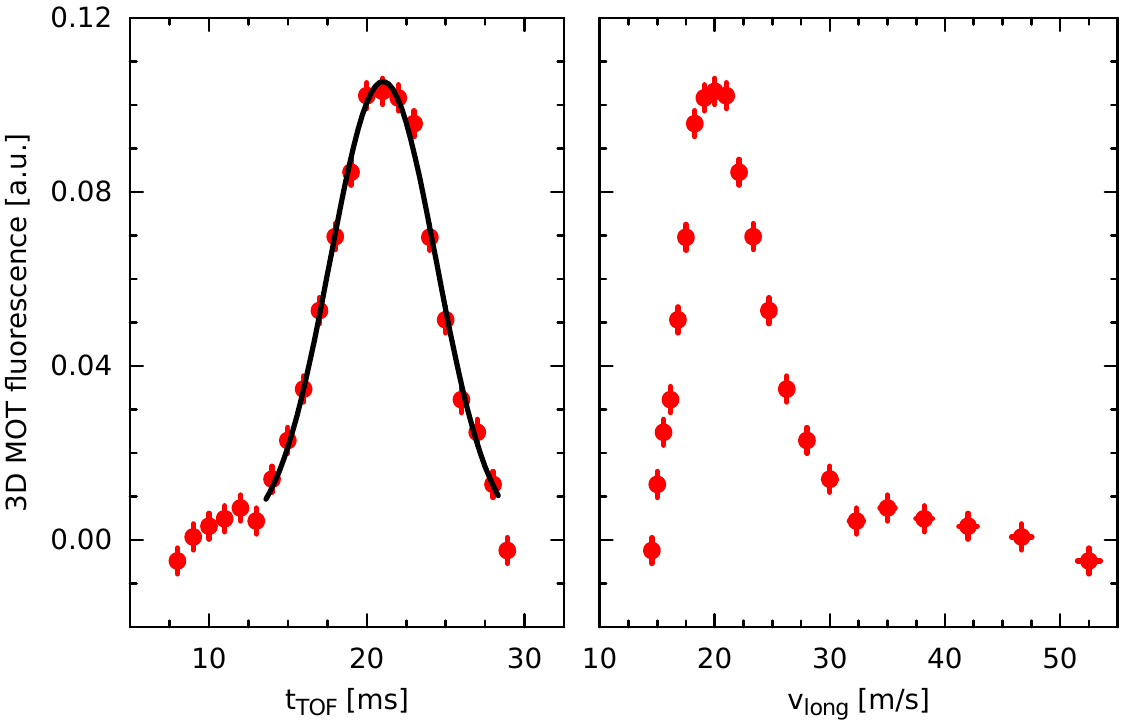}
\caption{(Color online) Measured time of flight distribution of the atomic beam (left). Velocity distribution (right) of the atomic beam deduced from the time of flight and the known distance between the MOTs.}
\label{fig:TOF}
\end{figure}

The longitudinal velocity distribution of the atomic beam from the \mott was measured with a time-of-flight (TOF) method. The population of a specific velocity class of the source can be measured with the following procedure. Starting from an empty \mottt without the push beam, a 1~ms long pulse of the push beam launches an atomic packet from the \mott while the \mottt repumping light is switched off. After a given time $t_{\text{TOF}}$ the repumping light is switched back on for 3~ms time, so that only atoms traveling with a velocity $v_{\text{long}}=\frac{d_{\text{TOF}}}{t_{\text{TOF}}}$ ($d_{\text{TOF}}$=42$\pm1$~cm is the distance between \mott and \mottt centers) are recaptured in the \mottt. In order to increase the signal-to-noise ratio this procedure is repeated 450 times every 34 ms, accumulating atoms in the trap, before acquiring the signal reported in Fig.~\ref{fig:TOF}.

The distribution of $t_{\text{TOF}}$ was fitted to a Gaussian function finding a mean value of $t_0$=$(21.0\pm0.1)$~ms and a standard deviation $\Delta t$=$(3.4\pm0.1)$~ms (Fig.~\ref{fig:TOF} left). The velocity distribution (Fig.~\ref{fig:TOF} right) is therefore centered at $v_0$=$(20.0\pm0.5)$~m/s and has a half-width $\Delta v_{\text{long}}$ of the order of 3~m/s. This measurement of the peak velocity is not limited by the recapture velocity of the \mottt, expected to be much larger, nor by the gravity fall during the transfer, that introduces a lower cut-off at 2~m/s, considering our geometry. \

The residual longitudinal temperature of the atomic beam can be estimated from the width of the velocity distribution as $T_{\text{long}}$=$m\,\Delta v_{\text{long}}^2/k_{\text{B}}\sim 30$~mK. In the transverse direction, one can consider the solid angle covered by the \mottt capture area and the distance $d_{\text{TOF}}$, obtaining $1.8\times10^{-3}$~sr. The measured atomic flux of $4.5\times10^9$~atoms/s implies then a source brightness larger than $2.5\times10^{12}$ atoms/s/sr. The differential pumping  channel sets a geometrical upper limit on the velocity dispersion $\Delta v_{\text{trans}}$=0.9~m/s. This corresponds to a transverse temperature of the order of $T_{\text{trans}}$=2~mK, which amounts to a few times the Doppler temperature $T_{\text{D}}$=235~$\mu$K. The divergence set by the channel ($5.9\times10^{-3}$~sr) is larger than the measured one, therefore not limiting the atom transfer. \

The best performances of the atomic source are summarized in Table~\ref{tab:max-flux}. From the values reported in that table we can see that the \zs beam increases the loading rate by a factor 12. This value is of the same order of magnitude of the factor estimated in the simulation described in the following section. \

\begin{table}[h]
\caption{Atomic source performances.}
\label{tab:max-flux}
\centering
\begin{tabular}{lll}
\toprule
loading rate without \zs    &\hspace{0.2 cm}   & $(3.7 \pm 0.5)\times10^8$~atoms/s       \\
loading rate with \zs       &   & $(4.5 \pm 0.5)\times 10^9$~atoms/s      \\
                                                                    \\
total trapped atoms         &   &                                   \\
in the 3D dark spot MOT      &   & $(1.2 \pm 0.1)\times 10^{10}$~atoms     \\
                                                                    \\
most probable         &   &                                   \\
longitudinal velocity      &   & $(20.0 \pm 0.5)$~m/s                  \\
                                                                    \\
source brightness      &   & $>2.5\times10^{12}$~atoms/s/sr                 \\
\hline
\hline
\end{tabular}

\end{table}

\section{SIMULATION OF \zs AND \mott OPERATION}   
\label{SIMULATION}
Slowing and cooling effects due to radiation pressure in presence of the magnetic field generated by the permanent magnets were simulated both for the first slowing stage and for the \mott. \

The numerical model is based on several simplifications: we restrict the simulation to a single two-level-atom trajectory in a 1D geometry, subject to classical radiation pressure. \

\begin{figure}[b]
\centering
\includegraphics[width=1\columnwidth]{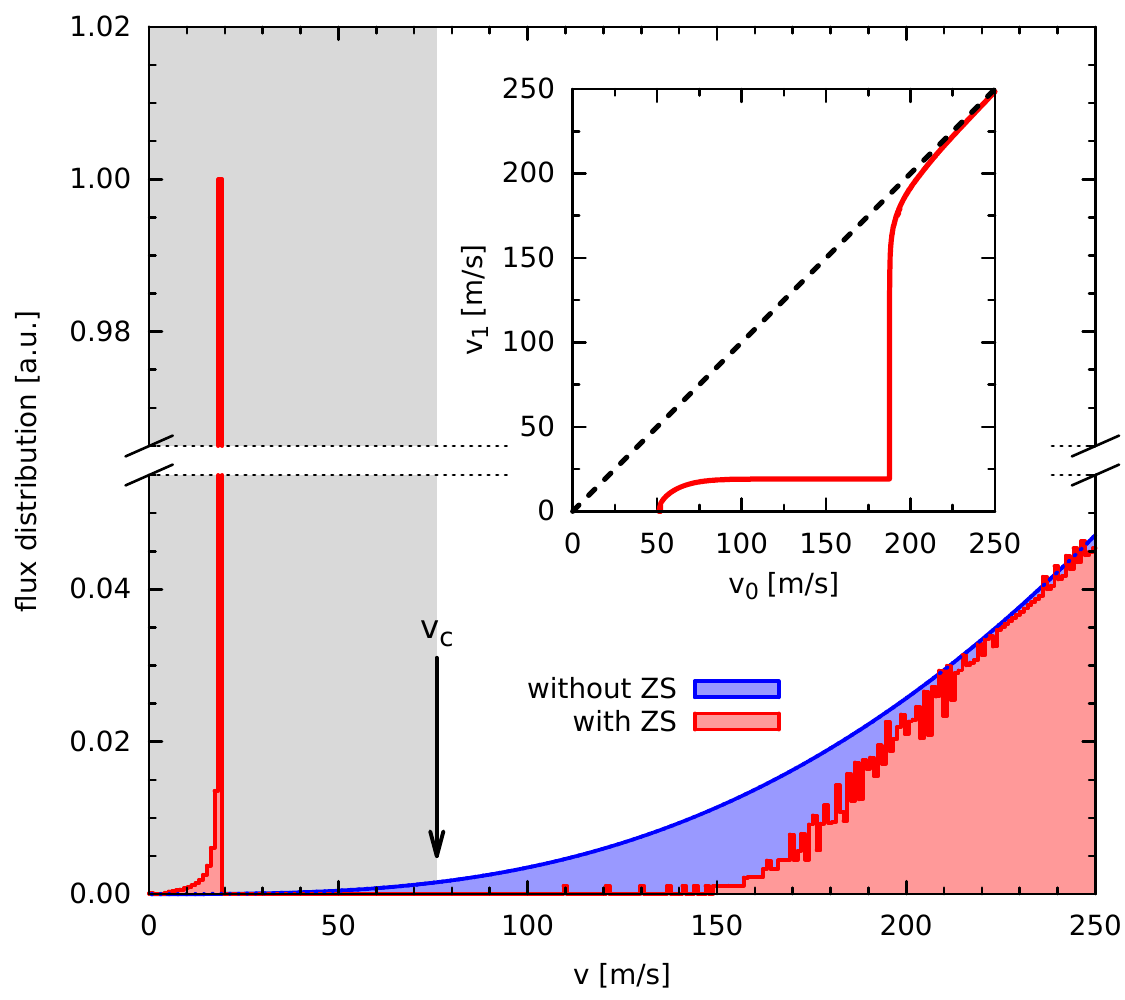}
\caption{(Color online) Flux distribution $\phi(v)$ of atoms emitted by the oven with (red) and without (blue) the \zs effect for a sample at 210$^{\circ}$C. Inset: \zs output velocity, $v_1$, as a function of the starting velocity at the oven, $v_0$. A large velocity class is slowed down to 19 m/s, below the \mott capture velocity $v_{\text{c}}$.}
\label{fig:simulation}
\end{figure}

The goal is to slow down the largest amount of atoms coming from the oven, below the \mott capture velocity $v_{\text{c}}$. We set the origin $z$=0 on the \mott axis and treat separately two regions: the \zs stage ($-126\text{~mm}<z<-17.7\text{~mm}$), and the \mott region ($-17.7\text{~mm}<z<17.7\text{~mm}$), the latter being delimited by the overlapping region between the \mott light beams. Since the \zs and the \mott beams have frequency offsets of several natural linewidths, we can argue that the two forces cannot be simultaneously active on atoms, thus justifying our separate treatment. The magnetic field is numerically calculated for the permanent magnets configuration described in Sec.~\ref{Magnetic field sources}. \

The starting conditions are the oven vertical position $z_0$=$-126$~mm and the initial velocity $v_0$, whereas $v_1$ represents the velocity at the end of the \zs region. For the simulations we use the parameters that have been experimentally optimized (see Tab.~\ref{tab:parameters}). \

Under the above conditions the atoms in the simulations are either bounced back by radiation pressure before reaching the \mott or not decelerated at all, depending on $v_0$. This is in contrast with the experimental evidence of gain when using the \zs. This simplified model serves as a demonstration of the system working principle, not as a quantitative calculation. We verified, though, that by considering a 20\% smaller magnetic field the simulation provides reasonable numbers in terms of efficiency of the cooling system, of the same order of the magnitude of the one experimentally observed. Moreover, efficiency and typical velocity classes involved seem not to change significantly as field or intensity parameters are slightly changed. Here we present the results we obtained by considering a 20\% smaller magnetic field, without further corrections on beams intensity. \

Atoms with $v_0$ ranging between 51~m/s and 187~m/s are decelerated to $v_1$=19~m/s, as shown in the inset of Fig.~\ref{fig:simulation}. For lower velocities the atoms are bounced back, while for higher velocities the \zs does not significantly affect the atomic trajectories. \

With similar methods we simulated the atomic trajectory in the \mott, taking into account the proper beams geometry and the magnetic field.
The initial position is now $z_1$=$-17.7$~mm. The atoms are trapped on a ms timescale if their initial velocity is smaller than the capture velocity, that is found to be $v_{\text{c}}$=76~m/s. This velocity is larger than the velocity of the atoms slowed down by the \zs stage. All the atoms involved in the \zs are thus captured in the \mott. \

The compact \zs stage is essential to change the velocity range of atoms involved in the full cooling process from 0-76~m/s to 51-187~m/s. This means more than just a 2 times larger velocity class because the atomic flux, considering the Maxwell-Boltzmann distribution, is not linear in $v$. Considering the Maxwell-Boltzmann distribution of a 1D thermal beam modified by the \zs (inset of Fig.~\ref{fig:simulation}), we calculate the atomic flux distribution $\phi(v)$ at the entrance of the \mott region (Fig.~\ref{fig:simulation}), which is substantially increased in the \mott capture range. The atomic flux distribution over an area $\sigma$ is $\phi(v)\,dv=n\sigma v\,f(v)\,dv$, where $n$ is the atomic density. \

We can integrate $\phi(v)$ from 0 to $v_{\text{c}}$ with and without the \zs. The ratio between the two integrals gives 35. This represents the gain factor given by the compact \zs stage to be compared with the experimentally measured flux ratio of 12. \

We are aware that the model is oversimplified and, in order to get an accurate quantitative result, one should have a better knowledge of the actual magnetic field \cite{Comment:field}, properly consider the multi-level atomic structure interacting with $\sigma^+$ and $\sigma^-$ light and extend the single axis model to a 3D trajectory. This goes beyond our goals. \

The numerical model can be adapted also for different simulations. In particular it was useful to predict whether with our setup the \mott would work correctly also for potassium, when the experiment will proceed towards the production of atomic mixtures. Trying different plausible parameters we found that the \mott can also trap all the stable isotopes of potassium. \

\section{CONCLUSION}     

In conclusion we reported on a new reliable scheme for producing a cold beam of sodium atoms which is characterized by compactness and high-flux performances at the same time. We demonstrated the possibility to efficiently trap sodium atoms in a \mott from a nearby, in-plane oven by first slowing them down within a short length in a \zs-like configuration. The design allows to introduce additional atomic species to cool with analogous procedure in the same system for the realization of multi-species experiments. \

The best loading rate in a 3D dark-spot MOT, located about 42 cm away from the source, of more than $4\times10^9$ atoms/s, allowed us to readily create a \mottt containing more than $10^{10}$ atoms. We observed that a few ms optical molasses stage \cite{Lett88,Comment:molasses} after loading a dark-spot MOT, consistently enhanced the phase space density even in the regime of high optical depth, similarly to what has been observed on $^{39}$K \cite{Landini11}. A large amount of atoms was thus transferred in a Ioffe-Pritchard \cite{Pritchard83} magnetic trap with axial and radial frequencies of 12 Hz and 128 Hz and performed evaporative cooling achieving a pure BEC with more than $10^7$ sodium atoms. \

We believe this novel and compact atomic source represents a valid alternative for the realization of a high-flux source of sodium atoms, especially in case of multi-species experiments where complexity and encumbrance are important issues. \

\section*{ACKNOWLEDGMENTS}
The authors thank R. Graziola, M. Tomasi and the Electronic\&Engineering Service of the University of Trento for the valuable and effective support provided during the activation of our laboratory. We also acknowledge fruitful discussions with M. Prevedelli and F. Schreck, the support of the whole BEC Center in Trento and the collaborative exchange with the Quantum Degenerate Gases group at LENS in Florence. This work was financially supported by Provincia Autonoma di Trento.

\end{document}